# Unlocking the hybrid piezo and pyroelectric nanogenerators performance by SiO$_2$ nanowires confinement in poly(vinylidene fluoride)

Juan Delgado-Alvarez,[1] Hari Krishna Mishra,[1]* Francisco J. Aparicio,[1] Xabier García-Casas,[1] Angel Barranco,[1] Juan R. Sanchez-Valencia,[1] Victor Lopez-Flores,[1,2]* Ana Borras[1]

1. Nanotechnology on Surfaces and Plasma Laboratory, Materials Science Institute of Seville, CSIC-US, Avd. Americo Vespucio 49, 41092, Seville, Spain
2. Departamento de Física Aplicada I, Escuela Politécnica Superior, Universidad de Sevilla, Virgen de Africa, 41011, Seville, Spain

**Abstract**
We report on the development of a novel flexible piezo/pyro-electric nanogenerator (PPNG) that combines a uniform film of poly(vinylidene fluoride) (PVDF) infiltrated over vertically supported SiO$_2$ nanowires (NWs) to enhance both piezoelectric and pyroelectric energy harvesting capabilities. The synthetic procedure involves a low-temperature multi-step approach, including the soft-template formation of SiO$_2$ NWs on a flexible substrate, followed by the infiltration of a PVDF thin film (TF). The plasma-enabled fabrication of SiO$_2$ NWs facilitated vertical alignment and precise control over the surface microstructure, density, and thickness of the confined nanostructures. These strategic structural systems promote the development of the most favourable electroactive β- and γ-phases in the PVDF matrix. Notably, the electrical poling plays a major role in aligning the random dipoles of the PVDF macromolecular chain in a more ordered fashion to nucleate the amplified electroactive phases. As a proof-of-concept, the fabricated PPNG exhibited a significant improvement in the instantaneous piezoelectric output power density (P), ~ 9-fold amplification relative to its bare PVDF TF counterpart. Analogously, the pyroelectric coefficient (p) demonstrated a 4-fold superior performance with referenced PVDF TF based PPNG. Thus, the engineered system of SiO$_2$ NWs@PVDF comprising PPNG offers a promising pathway toward multisource energy harvesting capabilities through efficient energy transduction at mechanical excitation frequencies of 10-12 Hz and across a temperature difference (ΔT) of 9 to 22 K.

## Introduction

The growing global demand for sustainable and miniaturized electronic devices has prompted extensive research into alternative sources of energy harvesting technologies.[1–3] Particularly, the development of smart self-powered electronic systems represents a pivotal strategy for mitigating dependence on external power sources within electronic industrialization, thereby addressing critical energy and environmental concerns.[4] Of particular interest, nanogenerators (NGs) that convert ambient untapped mechanical and thermal energies of low frequencies into usable electrical power are increasingly gaining attention for low-power electronics.[5–7] For instance, piezoelectricity refers to the variation of spontaneous polarization in response to tunable mechanical stimuli, whereas pyroelectricity describes the change in spontaneous polarization as a function of temperature oscillations.[8–11] NGs, which rely on piezo,- pyro,- and tribo-electric mechanisms, offer a compelling solution for powering low-energy applications such as sensors, wearables, and medical devices.[12,13] Notably, the development of efficient hybrid and multisource nanogenerators capable of simultaneously converting from vibrations and temperature variations could help to solve the major challenge in energy harvesting, i.e., the inherent random and intermittent nature of these ubiquitous environmental energy sources.[14,15]

In this regard, poly(vinylidene fluoride) (PVDF) and its copolymers have emerged as a leading material for flexible and biocompatible nanogenerators due to their remarkable piezo- and pyro- electric properties, mechanical flexibility, chemical stability, and transparency.[16] These characteristics make PVDF an ideal candidate for energy harvesting, sensing, ferroelectric memory devices, and biomedical applications.[17] However, PVDF-based devices, while effective, often suffer from limited energy output due to the semicrystalline nature of PVDF, which induces crystallization in non-electroactive phases.[18]

PVDF exhibits different polymorphic phases, such as α, β, γ and δ depending on the macromolecular chain's dipole ($-CH_2-$/$-CF_2-$) arrangements. Among them, α-phase exhibits an alternative trans (T)−gauche (G) stereochemical chain conformation, non-polar and thermodynamically stable, while the β-, γ-, and δ-phases are polar with parallel chain packaging of dipole orientation with non-zero dipole moment and therefore, electroactive.[19,20] Among these electroactive phases, the β-phase is generally preferred due to its highest electrical dipole moment and superior piezoelectric performance. The piezo- and pyro-electric properties of PVDF are mainly governed by the electroactive phases, which result from the crystallization of the PVDF macromolecular chain.[21] Notably, to achieve the desired electroactive phase in PVDF, the most commonly known copolymer of PVDF, i.e., poly(vinylidenefluoride-trifluoroethylene) (PVDF-TrFE), is often adopted. This copolymer tends to naturally crystallize in electroactive β-phase, eliminating the need for additional processing. However, the high cost of the copolymer and the lower value of the critical temperature have sparked interest in enhancing pure PVDF performance through composite materials and nanostructures.[22,23]

In addition, a promising approach to significantly improve the efficiency of PVDF nanogenerators involves the integration of functional nanofillers. In particular, the incorporation of metal oxide nanomaterials into PVDF matrices has been shown to substantially boost the piezoelectric output of these nanogenerators due to the preferential promotion of electroactive β- and γ-phases during PVDF crystallization, driven by oxide-polymer interactions.[24,25] Indeed, multiple aspects have been proposed to elucidate the positive role of nanofillers such as ZnO, $BaTiO_3$, $TiO_2$, $Co_3O_4$, $Fe_2O_3$, and $Al_2O_3$ nanoparticles and nanowires.[26–29] These mechanisms encompass: (i) Augmented interfacial interactions: which facilitate a more favourable alignment of electroactive PVDF chains. (ii) Reduced nucleation energy barrier: the nanofiller surfaces act as heterogeneous nucleation sites, lowering the energy required for the formation of the polar β- and γ-phases. (iii) Electrostatic interactions: arising from the inherent surface charges of the nanofillers and the induced mechanical stress inherent to the mismatch in the mechanical properties of the PVDF matrix and the metal oxide nanostructures.[22,26] Moreover, the nanofillers dispersion, maximizing surface contact interaction, plays a critical role in promoting the enhanced formation of the electroactive phases.[30] The most common nanofillers take the form of nanoparticles and electrospun nanofiber composites.[30,31] Both of these frequently suffer from agglomeration, which reduces the available surface area of the nanofiller. Hence, a considerable effort focuses on developing strategies to counteract these shortcomings, such as the conformal formation of coatings or the inclusion of additives to provide uniform dispersion.[32]

In this context, $SiO_2$ nanoparticles and one-dimensional nanostructures of MWCNTs have been previously integrated as nanofillers in PVDF matrix with enhanced β-phase to promote high X-band microwave shielding.[33] However, despite several efforts to enhance PVDF-based NGs, a systematic understanding of how nanoscale confinement influences energy harvesting performance remains limited.[34] Overcoming the agglomeration challenge is paramount for realizing the full potential of nanofiller-enhanced PVDF NGs. Thus, the major aim of this article is to demonstrate the enhanced performance of PVDF comprising confined $SiO_2$ nanowires (NWs) as a piezo/pyro-electric nanogenerator (PPNG). Our approach allows the formation of $SiO_2$ NWs on flexible supports as a guiding template by combining soft-template single-crystalline organic nanowires (ONWs) with plasma-enhanced chemical vapor deposition (PECVD) of $SiO_2$ shells. The density of these 1D nanostructures can be controlled to provide high open space to be filled by the PVDF, reducing agglomeration and promoting an enhanced interface area between the confined nanowires and the electroactive polymer (see schematic illustration of Fig. 1). The surface chemistry of $SiO_2$ is particularly important for its interactions with polymers. The surface of silica typically contains silanol groups (Si–OH), whose concentration is enhanced in $SiO_2$ films deposited by plasma-enhanced chemical vapor deposition (PECVD) at low or room temperature, as well as in plasma-treated $SiO_2$ surfaces, thereby increasing reactivity, hydrophilicity, and surface functionalization.[35,36] These silanol groups are highly reactive and play a pivotal role in mediating interactions with dipoles of PVDF.[37–41] The ability to functionalize these surface groups allows for tailored interactions, which is essential for controlling the properties of $SiO_2$@PVDF systems. We present the superior figures of merit of the confined system in comparison with standard PVDF TF comprising NGs and hypothesize that the dipole-alignment mechanism is driven by the vertically aligned dielectric nanostructures and effective alignment with electrical poling conditions. It is also noteworthy that, to the best of our knowledge, none of the proposed processing approaches so far offer the advantage of vertical alignment of the $SiO_2$ nanostructure as a template, enabling direct on-device synthesis.

In this work, we introduced the reliable growth of vertically aligned 1D $SiO_2$ NWs templates via the combination of ONWs soft templates and PECVD, subsequently confined them with a PVDF polymer thin layer. Upon electrical poling, this confined system has demonstrated a significant enhancement in the electroactive phase fractions; F(β) ~ 41 % and F(γ) ~ 59 %, respectively. This structural transformation amplifies the piezoelectric and pyroelectric responses under mechanical and thermal activation. The nanostructured PPNGs were tested under mechanical and variable temperature stimuli, presenting a considerable enhancement in the output performances for piezo and pyro outputs with respect to the pristine PVDF PPNGs thin film counterpart.

## Results and discussion

Fig. 1 provides a schematic representation of the multi-step experimental procedure employed for the growth of hierarchical nanostructures, their confinement within a PVDF thin layer, and the subsequent device fabrication process. The procedure evolves from a previously reported one-reactor fabrication of supported core@shell nanostructures (see experimental details in the Experimental Methods section).[42–44] In this case, the process involves the deposition of an $SiO_2$ seed layer (Fig. 1a, step i), followed by the growth of organic $H_2Pc$ ONWs via physical vapor deposition (PVD) as shown in Fig. 1a, step ii), working as supported 1D templates. In the final step, a conformal $SiO_2$ shell layer is formed via PECVD to finally define the nanostructures (Fig. 1a, step iii). The formation of the shell by downstream PECVD allows for the straightforward control of the shell thickness and microstructure, and their vertical alignment is promoted by the plasma shield electric field.[42] Subsequently, a thin layer of PVDF polymer was introduced to confine these vertically aligned $SiO_2$ NWs (Fig. 1a, step iv). This confinement is crucial for influencing the crystallization behaviour of PVDF. Finally, the device architecture was completed by depositing top gold (Au) electrodes (Fig. 1a, step v) and establishing copper (Cu) strip wiring (Fig. 1a, step vi) to facilitate electrical measurements. In Fig. 1b, a comparison of the molecular structures of PVDF in its α-, γ-, and β-phases is presented, highlighting the significant dipole alignment observed in the β/γ-phase, which is essential for improved piezo- and pyroelectric performance. The marked domain provides an illustration of the unit cell for each phase, ordered to emphasize the significant attribute of a non-zero dipole moment in the β-and γ-polymorphs, a feature absent in the α-phase.

A plausible mechanism, visually elucidated in Fig. 1c, underpins the promotion of electroactive phases within this composite system through a dual interplay of surface chemistry. Firstly, hydrogen bonding plays a pivotal role: the surface of the $SiO_2$ NWs typically contains a high concentration of polar silanol (Si-OH) groups.[37–40] These groups are favourably enhanced in $SiO_2$ deposited by PECVD at room temperature.[35,45] Concurrently, the highly electronegative fluorine atoms within the PVDF polymer chains possess readily available lone pairs of electrons, rendering them potent hydrogen bond acceptors. This molecular complementarity facilitates the formation of robust hydrogen bonds between the fluorine atoms of PVDF and the hydrogen atoms of the silanol groups on the $SiO_2$ surface.[37,38] Such crucial anchoring intermolecular interactions effectively overcome the inherent conformational energy barrier that typically favours the non-polar α-phase of PVDF, thereby promoting the reorientation of PVDF chains into the highly polar all-trans (TTTT) conformation characteristic of the β-phase and TTTGTTT for γ-phase. Secondly, electrostatic interactions contribute significantly to the surface of the $SiO_2$ NWs, which inherently carry a negative charge.[46] Conversely, the -$CH_2$- groups within the PVDF polymer chains exhibit a partially positive charge density. This charge differential concludes in a compelling electrostatic attraction between the negatively charged $SiO_2$ surfaces and the positively charged -$CH_2$-dipoles of PVDF. This electrostatic interaction serves as an effective nucleation mechanism, inducing the precise alignment of PVDF chains into the all-trans conformation at the interface, thereby promoting the crucial transition to the polar and piezoelectric β- or γ-phases, which is a widely accepted phenomenon in PVDF nanocomposites.[33,47,48] The synergistic effect arising from the combined influence of both hydrogen bonding and electrostatic interactions profoundly contributes to the significantly enhanced β- and γ-phase content observed within the PVDF layer. The presence of interfacial interactions between the PVDF layer and the SiO$_2$ NWs suggests that SiO$_2$ facilitates the stabilization of the electroactive phase in PVDF. Fig. 1d demonstrates the dual-mode operation of the device: (i) in piezoelectric mode, where mechanical deformation in cantilever geometry generates voltage due to dipole realignment, and (ii–iii) pyroelectric mode, where heating or cooling induces electrical output in response to temperature stimulus. This multifunctional configuration enables the device to harness the potential for mechanical and temperature sensing for wearable electronics and environmental monitoring.

Fig. 2 shows representative SEM micrographs of the multi-step formation of the SiO$_2$ NWs@PVDF composite. The starting point of this process is the evolved one-reactor synthesis of core@shell nanostructures. This protocol combines the formation of 1D and 3D organic small-molecule single-crystalline nanowires (ONWs) with the conformal growth of metal oxide shells under mild temperature, vacuum, and power conditions enabled by plasma. Previous exploitation of this approach has allowed for the development of transparent conducting nanotubes and nanotrees,[44,49] ZnO piezoelectric core@shell nanowires[43,50] and highly stable metal halide perovskite nanotubes,[51] as well as SERS multifunctional sensors,[52,53] nanoscale waveguides,[42] superhydrophobic and anti-icing surfaces.[54] A significant advantage of this protocol involves the direct deposition of the organic soft-template NWs on processable substrates, requiring only the presence of nucleation centres, developing a certain surface roughness, and prompting the supersaturation regime needed for the formation of single-crystalline 1D nanostructures. Fig. 2a) shows the cross section of the SiO$_2$ employed as a seed layer (see Experimental Methods). Even the smooth surface of the ~150 nm layer allows for the formation of H$_2$Pc NWs, which appear as randomly oriented, growing from the SiO$_2$ surface. The conditions for the ONWs formation have been tuned to yield a length of 2-3 μm on average and a squared footprint between 50 and 150 nm (Fig. 2b). In the second step (see Schematic Fig. 1), the SiO$_2$ shell is deposited at room temperature in the down-stream region of a MW ECR (microwave electron cyclotron resonance) PECVD reactor. The selected plasma conditions provide both conformal formation of the SiO$_2$ shell and vertical alignment of the nanowires, as evidenced by the cross and top view micrographs in panels c and d, respectively. The formation of the shell induces the vertical alignment of the NWs by a combination of the action of the plasma shield electric field, perpendicular to the substrate, and enhances the stiffness of the NWs by adding the SiO$_2$ shell. The shell develops a nanoporous globular microstructure and is conformal, covering as well the interface with the seed layer, a slightly thicker layer towards the tip due to self-shadowing effects inherent to the subsequent growth on the aligned NWs.[42] The density of NWs can be controlled by the seed layer roughness and PVD conditions for the growth of ONWs. In this case, we have selected conditions for a density in the range of 3-4 H$_2$Pc@SiO$_2$ NWs/μm$^2$. It is expected that the combination of open porosity between the NWs and the high surface area of the nanoporous conformal shells will facilitate the PVDF infiltration by spin-coating (as depicted in Fig. 1). Thus, the cross-section micrograph in panel 2f shows the complete coverage of the NWs by the PVDF thin layer, which presents a smooth morphology. It is worth noting that the microstructure of the PVDF is free from any signs of agglomeration or pinholes. Thus, the PVDF matrix fully permeates the SiO$_2$ NWs down to the substrate, maintaining a compact structure with a uniform thickness of around ~ 3 ± 0.25 μm. The high PVDF molecular weight gave a viscous solution that granted a thick

PVDF layer at the selected rotating speeds during the spin-coating process. This also added some over-the-top thickness to guarantee that the infiltrated PVDF layer covers their full length and thus prevents short-circuits between the top and bottom electrodes in the formation of the nanogenerators. Figs. 2e and 2f compare the cross-section of the reference PVDF TF and polymer-confined SiO$_2$ NWs system, confirming that the presence of the supported nanostructures does not alter the thickness or morphology of the polymer. These micrographs also show the entire architecture of the assembled nanogenerators, including the bottom and top electrodes, formed by ITO and Au layers as schematized in Fig. 1a.

It has been widely reported that the piezoelectric and pyroelectric performance of the electroactive phases of PVDF can be enhanced, apart from the use of nanofillers, by alignment of the PVDF chains upon electrical bias, strain, or by physical confinement.[55] Among them, electrical poling in different combinations of stretching, annealing, and pressure-induced crystallization has been exploited to transform from α-phase or direct formation of the β-phase.[56] In this article, we have taken advantage of the direct formation of the vertically aligned nanofillers on conductive substrates (Indium tin oxide coated polyethylene terephthalate (ITO/PET)) to carry out in-situ poling on the SiO$_2$ NWs@PVDF system by imposing an electric field between the top and bottom electrodes, while keeping the entire system at ambient temperature (see Fig. 1 and experimental details in Methods). Exhibiting notable advantages, the electrical poling method proves to be a remarkably compatible technique, readily adaptable to a wide spectrum of substrate materials.

Fourier Transform Infrared (FTIR) spectroscopy in ATR mode is carried out to analyse the effect of interfacial interaction between PVDF dipoles and SiO$_2$ NWs, and also how electrical poling at varying voltages can nucleate the most prominent electroactive phase in the system. ATR spectra in Fig. 3a show the normalized absorbance of different PVDF-based samples to visualize the characteristic features of PVDF polymorphs. It is noteworthy that the spectra were normalized to the intensity of 1071 cm$^{-1}$ vibrational bands due to their linear dependence on the film thickness, regardless of crystalline nature.[57] Specifically, the spectrum of the pristine PVDF thin film (labelled as PVDF TF) exhibits prominent peaks at 838 cm$^{-1}$, attributed to overlapping β- and γ-phases with almost no α-phase peak at 763 cm$^{-1}$. However, the most prominent signature of γ-phase is noticed in PVDF TF due to the presence of intense bands at 812 and 1234 cm$^{-1}$, respectively.[57] Further, the incorporation of SiO$_2$ NWs into the PVDF matrix (labelled as SiO$_2$ NWs@PVDF-UP for unpoled) shows a noticeable change in the relative intensities of these characteristic peaks at 812, 840, and 1234 cm$^{-1}$. These intensities increase considerably compared to those in pristine PVDF TF, suggesting a potential enhancement in the electroactive β and/or γ phases upon the introduction of the SiO$_2$ NWs confinement within the PVDF matrix. It is important to note that, although no signature of the non-polar electroactive α-phase of PVDF is observed in the composite system while the enhanced γ-phase is prominent, there is yet significant potential to increase the electroactive content of the aligned dipoles in the all-trans conformation through electrical poling. Fig. 3a also reveals the significantly enhanced electroactive β-phase fraction upon electrical poling at tunable voltage ranges from 100-200 V. The poling was carried out in-device, i.e., by applying the voltage across the top and bottom electrodes, ensuring high homogeneity along the entire device. In the normalized spectra of poled SiO$_2$ NWs@PVDF (labelled as SiO$_2$

NWs@PVDF_P), the characteristic vibrational peak of β-phase of PVDF at 1276 cm$^{-1}$ becomes more pronounced along with a substantial rise in γ-phase signature with the electrical poling voltage. To obtain a clearer visualization, an enlarged view is displayed in Fig. 3b, marked with colours, showing the signature band of β- and γ-phases at various poling voltages. Notably, it is clear from ATR spectra that electrical poling is sufficient to tune and align the dipoles of PVDF unidirectionally. Moreover, quantified electroactive phase fractions β- and γ-phases as F(β) and F(γ), respectively, were estimated by using the relations,[57]

$$F_{EA} = \frac{A_{840}}{\left(\frac{K_{840}}{K_{763}}\right)A_{763} + A_{840}} \times 100\ \% \tag{1}$$

$$F(\beta) = \frac{A_{1276}}{A_{1276}+A_{1234}} \times F_{EA} \tag{2}$$

$$F(\gamma) = \frac{A_{1234}}{A_{1276}+A_{1234}} \times F_{EA} \tag{3}$$

where $F_{EA}$ is the total electroactive phase fraction. The $A_{840}$ and $A_{763}$ imply the absorbance intensities at 840 and 763 cm$^{-1}$, respectively. $K_{840}$ (7.7×10$^4$ cm$^2$ mol$^{-1}$) and $K_{763}$ (6.1×10$^4$ cm$^2$ mol$^{-1}$) are the absorption coefficients corresponding to the respective wavenumbers. $A_{1276}$ and $A_{1234}$ indicate the intensities regarding the β- and γ- phases respective peaks. The quantitative analysis of PVDF phases reveals a clear correlation between electrical poling voltage and the phase composition of the SiO$_2$ NWs@PVDF system (shown in Fig. 3c). Increasing the poling voltage from 100 V to 200 V results in a significant increase in the electroactive β-phase fraction, accompanied by a corresponding decrease in the γ-phase fraction (as marked by light orange colour). This demonstrates the effectiveness of electrical poling in promoting the formation and alignment of the dipoles, i.e., F(β) → 41 % and F(γ) → 59 %, which is crucial for enhancing the piezoelectric and ferroelectric properties of the system. The observed phase transformation demonstrates the potential of controlled electrical poling to tailor the PVDF phase composition for specific applications.

**Performance of piezoelectric signal acquisition**
To investigate the mechanical energy harvesting performance, the piezoelectric response of pristine PVDF TF and SiO$_2$ NWs confined PVDF films based PPNG was tested under mechanical stimulation in cantilever mode by magnetic shaker at an optimal frequency of ~ 11 Hz, as presented in Fig. 4. Figs. 4a and 4c illustrate the generated open-circuit voltage (V$_{OC}$) as a function of time for unpoled (UP) and poled (P) PPNG. Notably, the poled PVDF TF exhibits a higher voltage output (peak-to-peak V$_{OC}$ = 100 mV) compared to its unpoled counterpart (peak-to-peak V$_{OC}$ = 80 mV), underscoring the crucial role of electrical poling in enhancing the piezoelectric response of PVDF by aligning the polar -CF$_2$- dipoles within the polymer matrix and promoting the formation of the electroactive phase, as corroborated by the ATR results in Fig. 3.

The electrical instantaneous peak power density (P) as a function of external load resistance for the pristine PVDF TF based PPNG, shown in Fig. 4b, indicates a maximum P ~ 0.61 µW/m$^2$ for the unpoled case and 1.12 µW/m$^2$ for the poled, both achieved at an optimal load resistance (R) ~ 3 MΩ. The power density was estimated by using the relation $\left(P = V^2_{OC}/(R \times A)\right)$, where $A$ is an effective electrode area, these values highlight the inherent piezoelectric properties and the limitations of pristine PVDF for high-output energy harvesting applications.

In contrast, a significant enhancement in electrical output is observed for confined $SiO_2$ NWs with the PVDF layer, as depicted in Figs. 4c and 4d. The unpoled $SiO_2$ NWs@PVDF_UP already exhibits a considerably higher voltage output (peak-to-peak $V_{OC}$ = 0.8 V), i.e., an order of magnitude compared to the unpoled pristine PVDF TF ($V_{OC}$ = 80 mV). This suggests that the nanoconfinement of $SiO_2$ NWs facilitates some degree of polar phase formation or enhances the mechanical stress transfer within the composite even without external poling. The poled $SiO_2$ NWs@PVDF_P based PPNG demonstrates a further substantial improvement in its piezoelectric performance. The generated voltage reaches a peak-to-peak value of $V_{OC}$ ~ 3 V, significantly higher than that of the poled pristine PVDF ($V_{OC}$ = 100 mV). Consequently, the maximum P for the poled $SiO_2$ NWs@PVDF_P system (Fig. 4d) reaches an impressive value of 10.8 μW/m² at an optimal load resistance of ~ 100 MΩ. This represents an order of magnitude increase in power generation compared to the pristine poled PVDF PPNG (P ~ 1.12 μW/m²). The observed value of P is almost comparable to values reported for multi-layered PVDF-based nanogenerators.[58]

The enhanced performance of the $SiO_2$ NWs@PVDF can be attributed to several factors. Firstly, the high aspect ratio and stiffness of the $SiO_2$ NWs likely act as effective stress concentrators within the PVDF matrix, leading to a more efficient conversion of mechanical energy into electrical energy. Secondly, the presence of interfacial interaction can promote the nucleation and growth of the polar phases during the nanoconfinement and subsequent poling process, as evidenced by the ATR analysis (Fig. 3), which showed an increased β-phase content in the $SiO_2$ NWs-confined PVDF films, particularly after poling. The shift in the optimal load resistance towards higher values for the $SiO_2$ NWs@PVDF systems compared to pristine PVDF suggests a higher internal impedance, which is expected with the presence of the insulating $SiO_2$ NWs within the polymer matrix. Thus, the synergistic effect of the $SiO_2$ NWs and electrical poling leads to a substantial increase in both the generated $V_{OC}$ and P, making this system a promising candidate for advanced flexible and wearable energy harvesting devices.

**Temperature-induced pyroelectric signal generation**

The pyroelectric thermal energy harvesting performance of PPNG is examined in response to temporal temperature oscillations at varying temperature differences (ΔT = 9 to 22 K). Figs. 5a and 5b illustrate the input temperature oscillation profiles and their corresponding first-order rates of change (dT/dt) over time, respectively. These controlled thermal stimuli, with temperature variations (ΔT = 9 to 22 K), were applied to both unpoled (UP) and poled (P) PPNG. The output short-circuits current ($I_{SC}$) generated by the pristine PVDF TF and the $SiO_2$ NWs@PVDF system in response to ΔT is shown in Figs. 5c and 5d, respectively. For the pristine PPNG (Fig. 5c), a sharp pyroelectric current is observed upon the application of the cyclic temperature oscillations. As expected, the poled PVDF TF exhibits a significantly higher current output compared to its unpoled counterpart at cyclic ΔT = 9 K. This enhancement is directly linked to the alignment of the polar phases within the PVDF matrix achieved through the poling process, maximizing the change in spontaneous polarization with temperature. Furthermore, a prominent rise in $I_{SC}$ is observed with poled PVDF TF, but with ΔT variation of 15 and 22 K, respectively (Fig. 5c), which clearly attributes the change in spontaneous polarization due to temperature fluctuations. Similarly, the confined system of $SiO_2$ NWs@PVDF (Fig. 5d) demonstrates a markedly superior pyroelectric response compared to the pristine PVDF TF based PPNG in both poled and unpoled states, under similar temperature modulations. Upon poling, the pyroelectric current output of the $SiO_2$ NWs@PVDF system is further dramatically enhanced, reaching peak values 4 times higher than those observed for the poled pristine PVDF TF at ΔT = 22 K. In addition to the increased peak current, the duration of the current response is significantly prolonged, in contrast to the sharp, short-lived signal exhibited by the pristine PVDF TF. This extended current output suggests a sustained charge release over time, which is highly advantageous for practical thermal energy harvesting applications. These findings underscore the critical role of $SiO_2$ NWs confinement within PVDF, not only in boosting the magnitude but also in modulating the temporal characteristics of the pyroelectric response.

A quantitative evaluation of the pyroelectric coefficients (p) of the pristine PVDF TF and the $SiO_2$ NWs@PVDF system, respectively, as a function of the applied temperature difference presented in Figs. 5e and 5f. The p is calculated using the formula $I_{SC} = pA(dT/dt)$, where $I_{SC}$ is the pyroelectric current and A is the active area of the device.[59] For the pristine PVDF TF (Fig. 5e), the p increases with increasing ΔT for the poled samples and is estimated to be p ~ 0.35 μC/m².K at ΔT~ 22 K, which is more than 3 times higher than unpoled PVDF TF PPNG. In contrast, the $SiO_2$ NWs@PVDF system (Fig. 5f) exhibits significantly higher p across all temperature variations and poling states. The unpoled $SiO_2$ NWs@PVDF shows a p ~ 0.40 μC/m².K, which is already higher than the maximum value achieved by the poled pristine PVDF TF. Upon poling, the p of the $SiO_2$ NWs@PVDF system is further amplified, reaching an impressive value of ~ 1.4 μC/m².K at ΔT~ 22 K. This represents a remarkable enhancement of 4 times compared to the poled pristine PVDF TF PPNG under the identical conditions. The measured pyroelectric coefficient (p) demonstrates a substantial improvement over the reported value for a well-established $KNbO_3$-PDMS composite system (p ~ 0.8 μC/m².K).[60] Moreover, the nanoconfined $SiO_2$@PVDF system also showed potential in comparison with state-of-the-art pyroelectrics such as electrospun nanofibers prepared under high electric fields to promote most electroactive β-phase nucleation in PVDF (p ~ 4 nC/m².K) and PVDF/MWCNT composite nanofibers (p ~ 60 nC/m².K).[61] The observed enhancement in the pyroelectric performance is attributed to the typical dielectric nature of the $SiO_2$ NWs might influence the thermal stress and strain distribution within the nanoconfined system under heat rate expansion and contraction, potentially leading to an enhanced change in polarization with temperature. Thus, the engineered system $SiO_2$ NWs@PVDF enables the multisource energy harvesting capabilities to effectively harvest mechanical and thermal energy into usable electrical energy.

**Conclusion**

This study demonstrates a facile and effective methodology for significantly enhancing the energy harvesting capabilities of a nanogenerator through the strategic integration and nanoconfinement of vertically aligned $SiO_2$ NWs template within the ferroelectric polymer PVDF. The low-temperature, plasma-enabled fabrication process of $SiO_2$ enables precise control over the morphology and preferential growth of the $SiO_2$ nanostructures, which promotes the formation of the electroactive β- and γ-phases in PVDF. This structural design not only optimizes mechanical and thermal energy transduction but also addresses critical challenges related to nanofiller dispersion and interfacial interactions. The nanoconfined $SiO_2$ NWs@PVDF system promotes the formation of electroactive β- and γ-phases. It is worth mentioning that electrical poling efficiently facilitates the alignment of randomly orientated dipoles in-plane and orthogonal to the main C-C chain backbone of PVDF. As a proof-of-concept, the PPNG exhibited a significant improvement in the instantaneous piezoelectric output power density (P) ~ 10.8 µW/m$^2$, a potentially enhanced value compared to ~ 1.2 µW/m$^2$ achieved by a bare PVDF TF based PPNG. Moreover, the PNNG has also demonstrated a substantial improvement in the pyroelectric coefficient (p) from the 0.35 µC/m$^2$.K observed for the bare PVDF TF up to ~1.40 µC/m$^2$.K for the NWs-based PPNG counterpart. In conclusion, the hybrid PPNG has demonstrated a 9-fold higher piezoelectric output power density and a 4-fold higher pyroelectric coefficient than bare PVDF TF based PPNG. These substantially improved piezoelectric and pyroelectric performances, together with the mild conditions employed for the device fabrication and poling, open the path for applications in self-powered devices, wearable technologies, and sustainable energy solutions. This work provides a versatile platform for the development of high-performance energy harvesters by exploiting the synergistic interplay between ferroelectric polymers and rationally designed 1D vertical $SiO_2$ nanostructures.

**Experimental section**

**Materials.** Polyethylene terephthalate sheets coated with indium-tin oxide (PET/ITO; surface resistivity ~ 60 Ω/sq) substrates were purchased from Sigma-Aldrich and washed with ethanol before use. These substrates were used to fabricate the final device; additionally, doped Si substrates were used to allow for material characterization, such as cross-sectional surface morphology. PVDF pellets were purchased from Sigma-Aldrich, with an average molecular weight of (M$w$) ~ 5,34,000 g/mol.

**One-reactor synthesis of $SiO_2$ nanowires by soft-template and plasma-enhanced chemical vapor deposition (PECVD)**

Following the step-by-step method depicted in the schematic of Fig. 1 and previous references,[42,43,62] a first $SiO_2$ seed layer of thickness ~ 150 nm was grown using plasma-enhanced chemical vapor deposition (PECVD) in a microwave (2.45 MHz) electron cyclotron resonance (ECR) reactor in a downstream configuration.[50] The base pressure of the reactor was below 10$^{-5}$ mbar. Chlorotrimethylsilane (ClTMS) served as the silicon oxide precursor, which was purchased from Sigma Aldrich. The precursor was introduced into the chamber via a dosing line with a mass flow controller and dispersed onto the substrates. The dosing line and the mass flow controller were heated to ~ 60 °C to prevent condensation on the tube walls. The microwave plasma source (SLAN, Plasma Consult GmbH) was operated at 600 W. The total pressure during deposition was $10^{-2}$ mbar of pure $O_2$. The $SiO_2$ film was fabricated at room temperature, up to a thickness of ~ 260 nm at a growth rate of 3 nm/min. This seed layer provides the necessary roughness to act as a template to grow the organic nanowires. Organic nanowires of metal-free phthalocyanine $H_2Pc$ were grown on the $SiO_2$ seeds by physical vapor deposition (PVD) (see Borrás et al).[63] The base pressure of the chamber was $10^{-6}$ mbar, and the working pressure was $10^{-2}$ mbar of Ar gas. The sample holder temperature was set to 190 °C, while the crucible temperature was kept at ~ 300 °C to maintain a constant growth rate of 0.45 Å/s as monitored by the quartz microbalance positioned near the substrate. The growth process was controlled to limit the total length of the $H_2Pc$ ONWs to ~ 2-3 μm, thereby forming the 1D soft templates for the subsequent $SiO_2$ shell. This $SiO_2$ shell was formed following the same PECVD described for the $SiO_2$ seed layer with a nominal diameter of 370 ± 30 nm.

As shown in Fig. 1, the PVDF thin film was infiltrated by spin-coating on the supported NWs to fill the free space among the NWs. The PVDF pellets were dissolved to a 15 wt% concentration in N, N-dimethylformamide (DMF) by continuously stirring the mixture for 24 h at 30 °C.[64] The resulting solution was then applied to the as-grown $SiO_2$ NWs via a two-step spin-coating: first, setting an initial rotation speed of 1500 rpm for 10 s, followed by an increase to 3000 rpm for 30 s; subsequently, the samples were immediately annealed on a hot plate at 60 °C for 2 h to ensure the complete solvent evaporation.

**Materials Characterizations**

The surface morphologies were obtained by scanning electron microscopy (SEM) on a Hitachi S4800, operating at 2 kV. Fourier transform infrared (FT-IR) spectra were registered on a JASCO FT/IR-6200 IRT-5000 in attenuated total reflectance (ATR) mode in the range from 1500 to 600 $cm^{-1}$ wavenumber, with 4 $cm^{-1}$ of spectral resolution. Electrical performance tests were conducted for fabricated PPNG before and after poling, using a Keithley 2635A source meter unit.

**Piezo/pyroelectric nanogenerator (PPNG) fabrication**

A section of the ITO/PET substrate was covered by a mask for the entire composite fabrication to ensure access to the bottom ITO electrode during device preparation. After the spin-coating and annealing of the PVDF, a Ti/Au (5/80 nm) top electrode was deposited onto the sample using thermal evaporation in a vacuum chamber with a base pressure of $10^{-6}$ mbar. During the evaporation process, the samples were maintained at a temperature of ~ 20 °C. The deposition rate was set to 0.5 Å/s using a quartz crystal microbalance (QCM). During this deposition, a second mask was added in order to deposit only onto the central part of the active area, avoiding short-circuits from the sides of the sample. Copper strips were affixed to the top and bottom electrodes using a silver paste to avoid direct use of the device electrodes during the performance tests, hence ensuring more robustness and longevity during its electrical testing (see the schematic of Fig. 1). Electrical poling was performed via the built-in top-bottom electrodes at ambient conditions for 30 min, with voltage varying between 100 and 200 V. It is worth mentioning that, a thin film sample (only PVDF TF) was fabricated with the identical conditions as above, except from the $SiO_2$ seed layer, the $H_2Pc$ nanowires and the $SiO_2$ coating, to serve as a reference without nanoconfinement.

In a cantilever configuration, piezoelectric signals were recorded from the PPNG fixed to a thin metallic scale driven at a defined frequency by a function generator for mechanical excitation (using magnetic smart shaker K2007E01). An optimal mechanical excitation frequency and amplitude were optimized to get the best output performance by the shaker. Pyroelectricity tests were executed by cyclic exposure of the top surface of the device to a hot air gun while refrigerating the bottom of the PPNG with a continuous cold air stream (see schematic in Fig. 1d). This method proved to deliver sharp temperature changes ($\Delta T$) in the ranges between 9 and 22 K with respect to ambient conditions while being non-invasive to the device. A thermocouple placed close to the device was linked to the electric measurements in order to ensure synchronization of the temperature measurements with output signals. The device was attached at the maximum bow angle position of the cantilever. The data analysis of electrical signals was carried out using the proprietary software NanoDataLyzer, available in open access.[65]

## ACKNOWLEDGMENTS
The authors thank the projects PID2022-143120OB-I00, TED2021-130916B-I00, and PCI2024-153451 funded by MCIN/AEI/10.13039/501100011033 and by "ERDF (FEDER)" A way of making Europe, Fondos NextgenerationEU and Plan de Recuperación, Transformación y Resiliencia". Project ANGSTROM was selected in the Joint Transnational Call 2023 of M-ERA.NET 3, which is an EU-funded network of about 49 funding organisations (Horizon 2020 grant agreement No 958174). The project leading to this article has received funding from the EU H2020 program under grant agreement 851929 (ERC Starting Grant 3DScavengers). V. L-.F. acknowledges the support of the VI Plan Propio de Investigación y Transferencia (VI PPIT) of the University of Seville.

## AUTHOR DECLARATIONS
**Conflict of Interest**
The authors have no conflicts to disclose related to this work.

**Author Contributions**
Ana Borras, Angel Barranco, Juan R. Sanchez-Valencia: conceptualisation and funding acquisition. Hari Krishna Mishra, Victor Lopez Flores and Ana Borras: writing original draft. Juan Delgado-Alvarez, Hari Krishna Mishra, Francisco J. Aparicio, Xabier García-Casas, Victor Lopez Flores: synthesis, characterization, device assembly. All authors were involved in the results investigation, methodology, validation, review, and editing.

## DATA AVAILABILITY
The data that support the findings of this study are available from the corresponding authors upon reasonable request.

**Corresponding Authors**
Email-Id: victor.lopez@csic.es and hari.krishna@icmse.csic.es

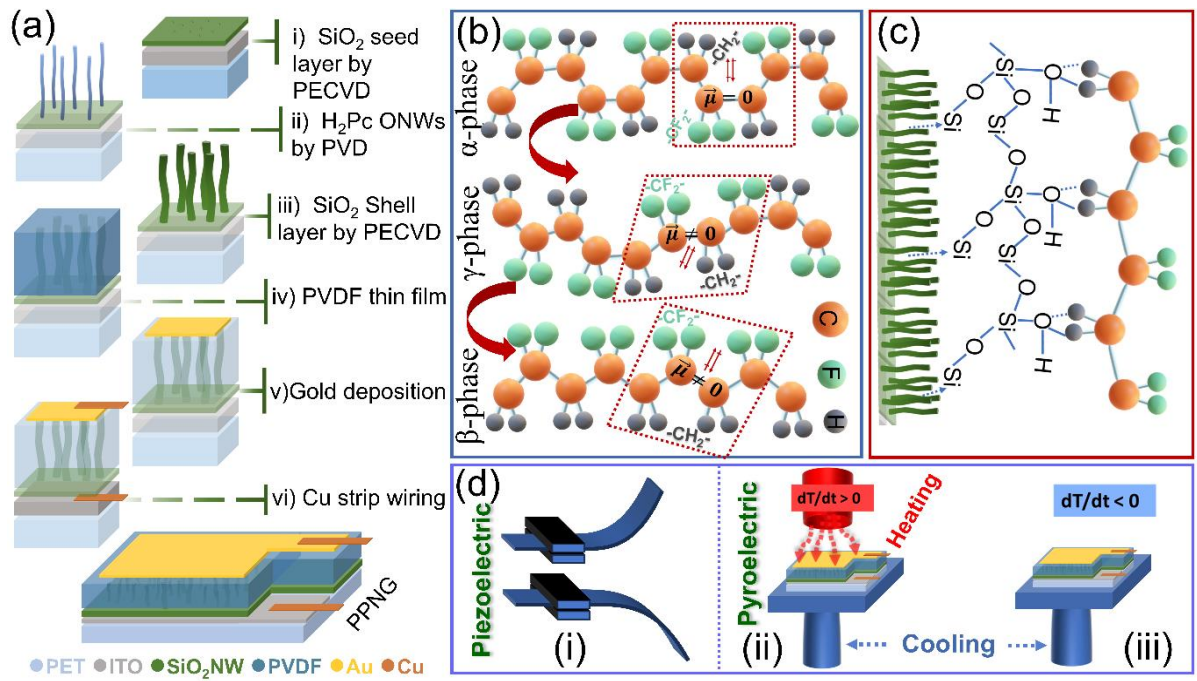

**FIG. 1.** (a) Schematic illustration of the fabrication process: vertically aligned SiO$_2$ nanowires (NWs) fabricated by a combination of ONWs soft-templates and conformal shell fabrication by PECVD, infiltration of PVDF thin film, and subsequent nanogenerator assembly. (b) Rearrangements of PVDF macromolecular chains under electrical poling, promoting nucleation of electroactive phases (β, γ). (c) Interaction between SiO$_2$ NWs and PVDF dipoles (–CH$_2$–/–CF$_2$– groups), facilitating dipolar alignment. (d) Experimental setup for nanogenerator characterization: (i) piezoelectric signal generation under mechanical excitation in cantilever mode; (ii) and (iii) pyroelectric response to thermal stimuli— heating and cooling, respectively.

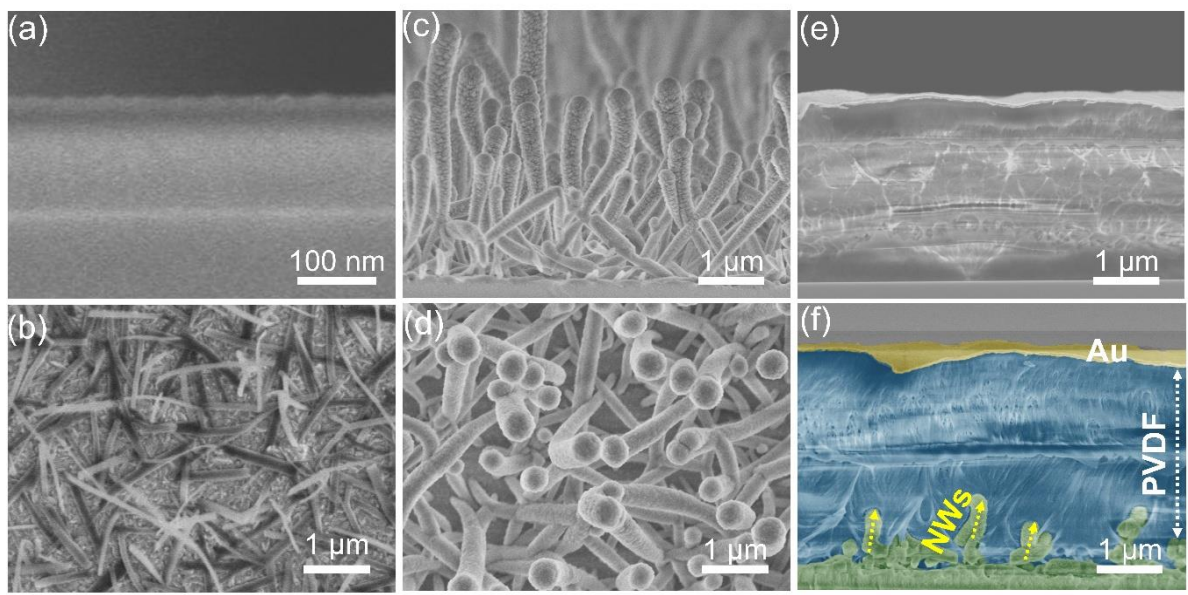

**FIG. 2.** Scanning electron microscopy (SEM) micrographs demonstrating key stages of the device architecture: (a) cross-sectional view of the SiO$_2$ seed layer thin film; (b) growth of H$_2$Pc organic nanowires (ONWs); (c) conformal growth of the SiO$_2$ shell on ONWs and preferential vertical alignment (cross-sectional view); (d) top-view of the SiO$_2$ shell morphology showing the open space between the nanowires; (e) cross-sectional morphology of the PVDF thin film (TF); (f) cross-sectional view of the final SiO$_2$ NWs@PVDF device architecture with top gold (Au) electrode and embedded NWs (as marked).

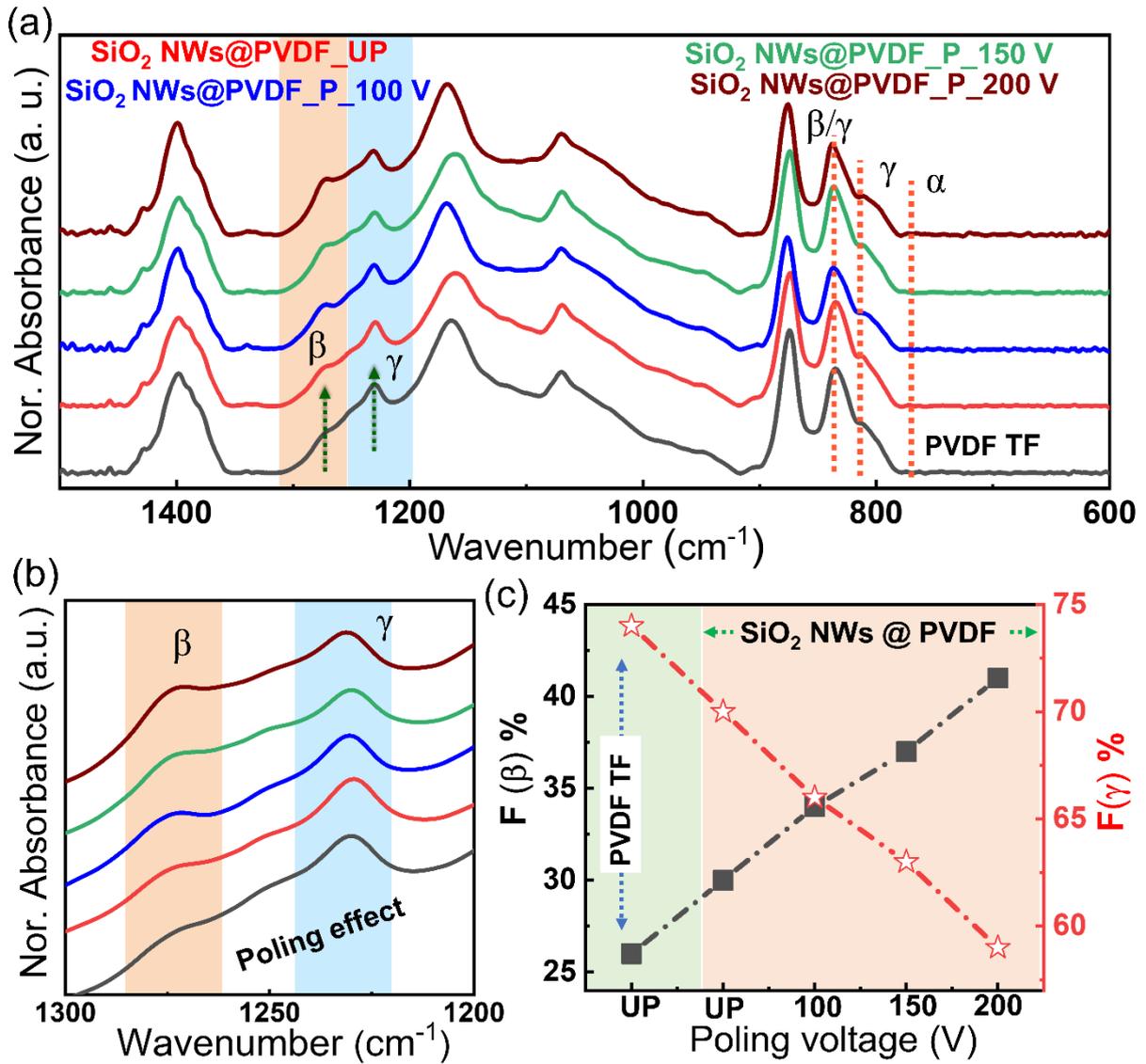

**FIG. 3.** (a) ATR-FTIR spectra of PVDF thin film (TF) infiltrated into the SiO$_2$ NWs-supported template, highlighting phase-specific vibrational features. (b) Magnified view of characteristic vibrational bands corresponding to the electroactive β- and γ-phases of PVDF. (c) Quantitative evolution of electroactive phase fractions, F(β) and F(γ), as a function of applied electrical poling.

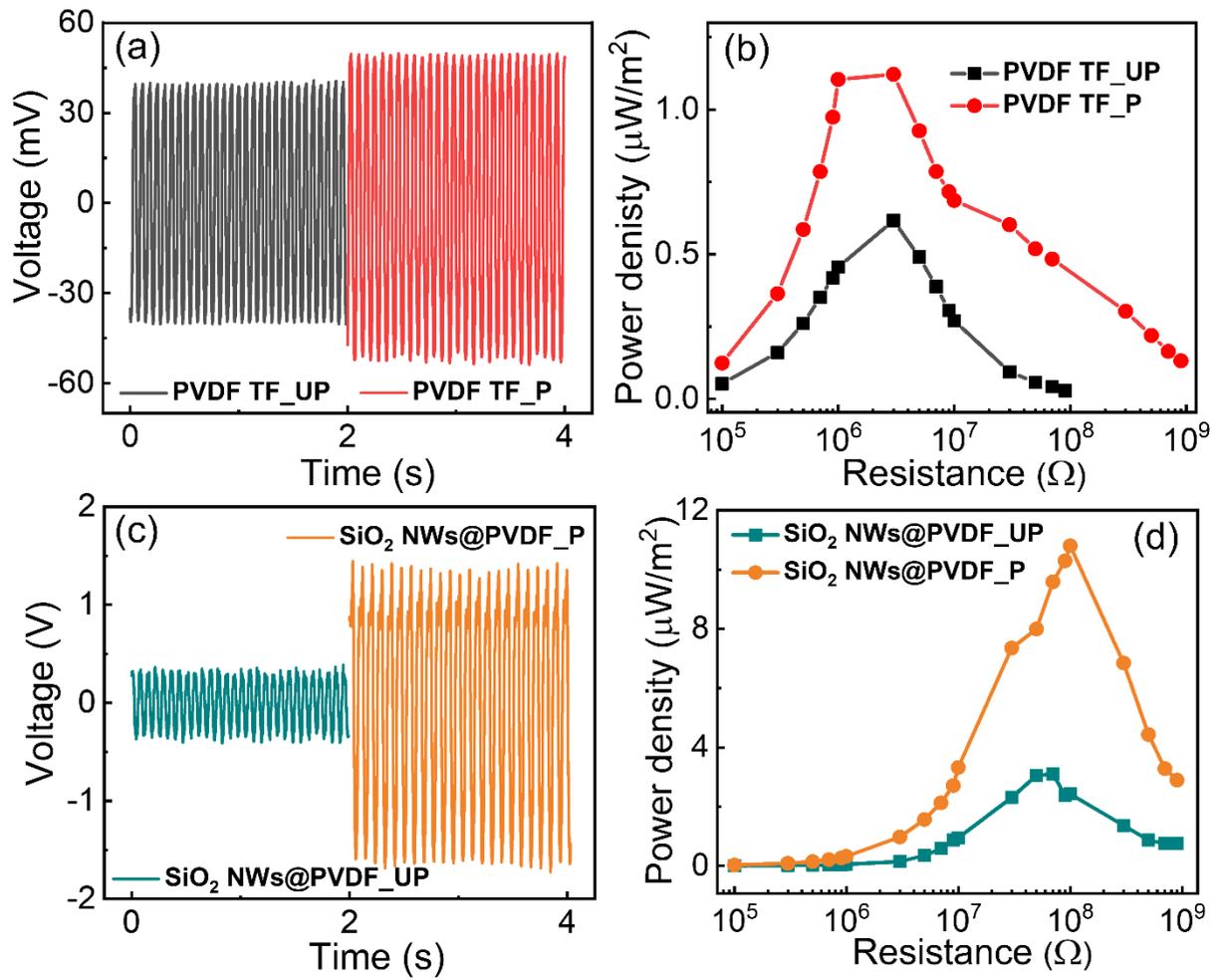

**FIG. 4.** Piezoelectric performance of the fabricated PPNG under mechanical excitation in a cantilever configuration. (a, c) Open-circuit voltage ($V_{OC}$) response of PVDF thin film (TF) and SiO$_2$ NWs@PVDF devices, respectively. (b, d) Corresponding instantaneous electrical peak power density (P) as a function of varying external load resistance.

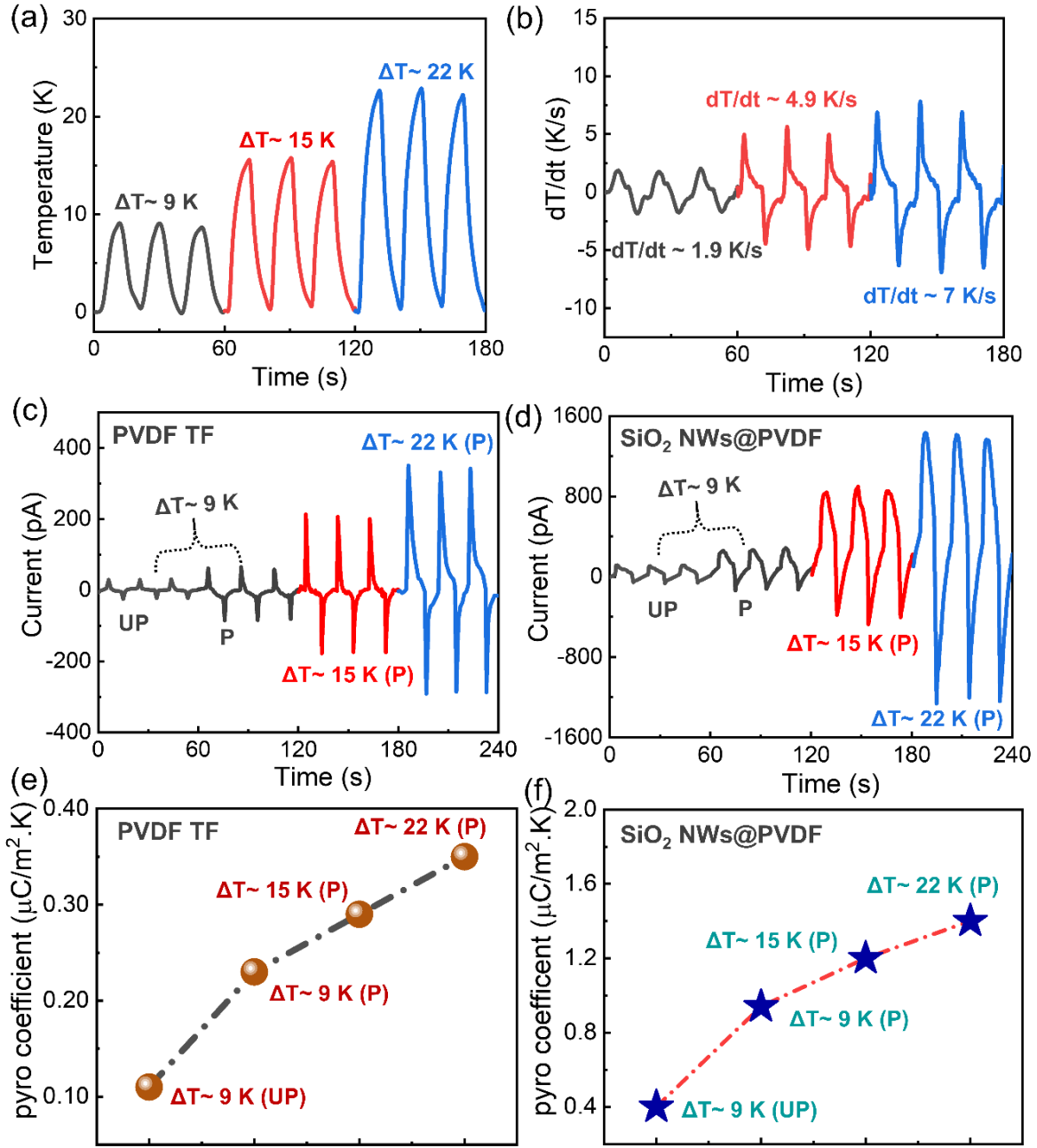

**FIG. 5.** Pyroelectric performance of fabricated PPNG under temperature stimuli. (a) Acquisition of the temperature oscillation profiles across varying temperature differences (ΔT → 9, 15, and 22 K). (b) Corresponding heat rate profile (dT/dt → 1.9 to 7 K/s). (c, d) Induced short-circuit pyroelectric current in PPNG in poled and unpoled PPNGs, respectively. (e, f) Extracted pyroelectric coefficients for both poled and unpoled PPNGs, highlighting the impact of poling and thermal response behaviour. The typical range of temperature variation was between 9 to 22 K with ambient conditions.

**References**

[1] G. Sim, H. Seo, J. Hwangbo, T. Kim, and Y. Choi, "Nano-structured piezoelectric polymers for biomedical application," APL Electron. Devices **1**(2), 021504 (2025).

[2] L. Xiao, B. Yin, Z. Geng, J. Li, R. Jia, and K. Zhang, "Flexible wearable devices based on self-powered energy supply," Nano Energy **142**, 111157 (2025).

[3] H. Li, C.R. Bowen, and Y. Yang, "Scavenging Energy Sources Using Ferroelectric Materials," Adv. Funct. Mater. **31**(25), 2100905 (2021).

[4] L. Guo, H. Wang, Z. Xu, R. Cong, L. Zhao, S. Zhang, K. Zhang, L. Gao, S. Wang, C. Pan, and Z. Yang, "Interfacial Pyro-Phototronic Effect: A Universal Approach for Enhancement of Self-Powered Photodetection Based on Perovskites with Centrosymmetry," Adv. Funct. Mater. **33**(49), 2306526 (2023).

[5] C.R. Bowen, H.A. Kim, P.M. Weaver, and S. Dunn, "Piezoelectric and ferroelectric materials and structures for energy harvesting applications," Energy Environ. Sci. **7**(1), 25–44 (2013).

[6] A.R. Jayakrishnan, J.P.B. Silva, K. Gwozdz, M.J.M. Gomes, R.L.Z. Hoye, and J.L. MacManus-Driscoll, "The ferro-pyro-phototronic effect for high-performance self-powered photodetectors," Nano Energy **118**, 108969 (2023).

[7] G. Velarde, S. Pandya, J. Karthik, D. Pesquera, and L.W. Martin, "Pyroelectric thin films—Past, present, and future," APL Mater. **9**(1), 010702 (2021).

[8] Z.L. Wang, "On Maxwell's displacement current for energy and sensors: the origin of nanogenerators," Mater. Today **20**(2), 74–82 (2017).

[9] P.K. Szewczyk, A. Gradys, S.K. Kim, L. Persano, M. Marzec, A. Kryshtal, T. Busolo, A. Toncelli, D. Pisignano, A. Bernasik, S. Kar-Narayan, P. Sajkiewicz, and U. Stachewicz, "Enhanced Piezoelectricity of Electrospun Polyvinylidene Fluoride Fibers for Energy Harvesting," ACS Appl. Mater. Interfaces **12**(11), 13575–13583 (2020).

[10] P.C. Sherrell, A. Šutka, M. Timusk, and A. Šutka, "Alternatives to Fluoropolymers for Motion-Based Energy Harvesting: Perspectives on Piezoelectricity, Triboelectricity, Ferroelectrets, and Flexoelectricity," Small **20**(32), 2311570 (2024).

[11] Y. Yang, W. Guo, K.C. Pradel, G. Zhu, Y. Zhou, Y. Zhang, Y. Hu, L. Lin, and Z.L. Wang, "Pyroelectric Nanogenerators for Harvesting Thermoelectric Energy," Nano Lett. **12**(6), 2833–2838 (2012).

[12] D. Saini, D. Sengupta, B. Mondal, H.K. Mishra, R. Ghosh, P.N. Vishwakarma, S. Ram, and D. Mandal, "A Spin-Charge-Regulated Self-Powered Nanogenerator for Simultaneous Pyro-Magneto-Electric Energy Harvesting," ACS Nano **18**(18), 11964–11977 (2024).

[13] "Paper-based ZnO self-powered sensors and nanogenerators by plasma technology," Nano Energy **114**, 108686 (2023).

[14] "Combination of power generation and flexibility in piezo-pyroelectric hybrid nanogenerators by designing bimetal textured structure," Nano Energy **116**, 108773 (2023).

[15] Y. Bai, H. Jantunen, and J. Juuti, "Energy Harvesting Research: The Road from Single Source to Multisource," Adv. Mater. **30**(34), 1707271 (2018).

[16] H.K. Mishra, A.K. Gill, V. Gupta, P. Malik, T.K. Sinha, D. Patra, and D. Mandal, "All Organic Aqueous Processable Piezo-Phototronic Ink for Strain Modulated Photoresponse," Adv. Mater. Technol. **8**(6), 2201350 (2023).

[17] M. Li, H.J. Wondergem, M.-J. Spijkman, K. Asadi, I. Katsouras, P.W.M. Blom, and D.M. de Leeuw, "Revisiting the δ-phase of poly(vinylidene fluoride) for solution-processed ferroelectric thin films," Nat. Mater. **12**(5), 433–438 (2013).

[18] A.J. Lovinger, "Ferroelectric Polymers," Science **220**(4602), 1115–1121 (1983).

[19] J. Martín, D. Zhao, T. Lenz, I. Katsouras, D.M. de Leeuw, and N. Stingelin, "Solid-state-processing of δ-PVDF," Mater. Horiz. **4**(3), 408–414 (2017).

[20] M. Kobayashi, K. Tashiro, and H. Tadokoro, "Molecular Vibrations of Three Crystal Forms of Poly(vinylidene fluoride)," Macromolecules **8**(2), 158–171 (1975).

[21] R.A. Surmenev, R.V. Chernozem, I.O. Pariy, and M.A. Surmeneva, "A review on piezo- and pyroelectric responses of flexible nano- and micropatterned polymer surfaces for biomedical sensing and energy harvesting applications," Nano Energy **79**, 105442 (2021).

[22] A. Gebrekrstos, T.S. Muzata, and S.S. Ray, "Nanoparticle-Enhanced β-Phase Formation in Electroactive PVDF Composites: A Review of Systems for Applications in Energy Harvesting, EMI Shielding, and Membrane Technology," ACS Appl. Nano Mater. **5**(6), 7632–7651 (2022).

[23] N. Meng, X. Zhu, R. Mao, M.J. Reece, and E. Bilotti, "Nanoscale interfacial electroactivity in PVDF/PVDF-TrFE blended films with enhanced dielectric and ferroelectric properties," J. Mater. Chem. C **5**(13), 3296–3305 (2017).

[24] Y. Su, W. Li, X. Cheng, Y. Zhou, S. Yang, X. Zhang, C. Chen, T. Yang, H. Pan, G. Xie, G. Chen, X. Zhao, X. Xiao, B. Li, H. Tai, Y. Jiang, L.-Q. Chen, F. Li, and J. Chen, "High-performance piezoelectric composites via β phase programming," Nat. Commun. **13**(1), 4867 (2022).

[25] X. Yuan, J. Shi, Y. Kang, J. Dong, Z. Pei, and X. Ji, "Piezoelectricity, Pyroelectricity, and Ferroelectricity in Biomaterials and Biomedical Applications," Adv. Mater. **36**(3), 2308726 (2024).

[26] S. Yempally, P. Magadia, and D. Ponnamma, "Effect of Zn–$Fe_2O_3$ nanomaterials on the phase separated morphologies of polyvinylidene fluoride piezoelectric nanogenerators," RSC Adv. **13**(48), 33863–33874 (2023).

[27] D. Zhang, X. Zhang, X. Li, H. Wang, X. Sang, G. Zhu, and Y. Yeung, "Enhanced piezoelectric performance of PVDF/$BiCl_3$/ZnO nanofiber-based piezoelectric nanogenerator," Eur. Polym. J. **166**, 110956 (2022).

[28] W.C. Gan, and W.H.A. Majid, "Effect of $TiO_2$ on enhanced pyroelectric activity of PVDF composite," Smart Mater. Struct. **23**(4), 045026 (2014).

[29] C. Wan, and C.R. Bowen, "Multiscale-structuring of polyvinylidene fluoride for energy harvesting: the impact of molecular-, micro- and macro-structure," J. Mater. Chem. A **5**(7), 3091–3128 (2017).

[30] B. Chai, K. Shi, Y. Wang, Y. Liu, F. Liu, L. Zhu, and X. Huang, "Integrated Piezoelectric/Pyroelectric Sensing from Organic–Inorganic Perovskite Nanocomposites," ACS Nano **18**(36), 25216–25225 (2024).

[31] M. Salama, A. Hamed, S. Noman, G. Magdy, N. Shehata, and I. Kandas, "Boosting piezoelectric properties of PVDF nanofibers via embedded graphene oxide nanosheets," Sci. Rep. **14**(1), 16484 (2024).

[32] C. Bahloul, S. Ez-Zahraoui, A. Eddiai, O. Cherkaoui, M. Mazraoui, F.-Z. Semlali, and M.E. Achaby, "Ferrite-doped rare-earth nanoparticles for enhanced β-phase formation in electroactive PVDF nanocomposites," RSC Adv. **14**(52), 38872–38887 (2024).

[33] E. Kar, N. Bose, B. Dutta, N. Mukherjee, and S. Mukherjee, "MWCNT@$SiO_2$ Heterogeneous Nanofiller-Based Polymer Composites: A Single Key to the High-Performance Piezoelectric Nanogenerator and X-band Microwave Shield," ACS Appl. Nano Mater. **1**(8), 4005–4018 (2018).

[34] V. Bhavanasi, D.Y. Kusuma, and P.S. Lee, "Polarization Orientation, Piezoelectricity, and Energy Harvesting Performance of Ferroelectric PVDF-TrFE Nanotubes Synthesized by Nanoconfinement," Adv. Energy Mater. **4**(16), 1400723 (2014).

[35] H. Ritala, J. Kiihamäki, and E. Puukilainen, "Correlation Between Film Properties and Anhydrous HF Vapor Etching Behavior of Silicon Oxide Deposited by CVD Methods," J. Electrochem. Soc. **158**(6), D399 (2011).

[36] R. Lukose, M. Lisker, F. Akhtar, M. Fraschke, T. Grabolla, A. Mai, and M. Lukosius, "Influence of plasma treatment on $SiO_2$/Si and $Si_3N_4$/Si substrates for large-scale transfer of graphene," Sci. Rep. **11**(1), 13111 (2021).

[37] D. Yuan, Z. Li, W. Thitsartarn, X. Fan, J. Sun, H. Li, and C. He, "β phase PVDF-hfp induced by mesoporous $SiO_2$ nanorods: synthesis and formation mechanism," J. Mater. Chem. C **3**(15), 3708–3713 (2015).

[38] Z. Yangzhou, Y. Weifeng, Z. Chaoyang, G. Bin, and H. Ning, "Piezoelectricity of nano-$SiO_2$/PVDF composite film," Mater. Res. Express **5**(10), 105506 (2018).

[39] E. Kar, N. Bose, S. Das, N. Mukherjee, and S. Mukherjee, "Enhancement of electroactive β phase crystallization and dielectric constant of PVDF by incorporating $GeO_2$ and $SiO_2$ nanoparticles," Phys. Chem. Chem. Phys. **17**(35), 22784–22798 (2015).

[40] J. Link, "Stabilization and structural study of new nanocomposite materials," (n.d.).

[41] H. Sun, Y. Liu, D. Li, B. Liu, and J. Yao, "Hydrophobic $SiO_2$ nanoparticle-induced polyvinylidene fluoride crystal phase inversion to enhance permeability of thin film composite membrane," J. Applied Polymer Science **136**(45), 48204 (2019).

[42] M. Macias-Montero, A.N. Filippin, Z. Saghi, F.J. Aparicio, A. Barranco, J.P. Espinos, F. Frutos, A.R. Gonzalez-Elipe, and A. Borras, "Vertically Aligned Hybrid Core/Shell Semiconductor Nanowires for Photonics Applications," Adv. Funct. Mater. **23**(48), 5981–5989 (2013).

[43] A.N. Filippin, J.R. Sanchez-Valencia, X. Garcia-Casas, V. Lopez-Flores, M. Macias-Montero, F. Frutos, A. Barranco, and A. Borras, "3D core-multishell piezoelectric nanogenerators," Nano Energy **58**, 476–483 (2019).

[44] J. Castillo-Seoane, J. Gil-Rostra, V. López-Flores, G. Lozano, F.J. Ferrer, J. P. Espinós, Kostya (Ken) Ostrikov, F. Yubero, A. R. González-Elipe, Á. Barranco, J. R. Sánchez-Valencia, and A. Borrás, "One-reactor vacuum and plasma synthesis of transparent conducting oxide nanotubes and nanotrees: from single wire conductivity to ultra-broadband perfect absorbers in the NIR," Nanoscale **13**(32), 13882–13895 (2021).

[45] A. Barranco, J. Cotrino, F. Yubero, J.P. Espinós, J. Benítez, C. Clerc, and A.R. González-Elipe, "Synthesis of $SiO_2$ and $SiO_xC_yH_z$ thin films by microwave plasma CVD," Thin Solid Films **401**(1), 150–158 (2001).

[46] F.M. Fowkes, and T.E. Burgess, "Electric fields at the surface and interface of $SiO_2$ films on silicon," Surf. Sci. **13**(1), 184–195 (1969).

[47] P. Martins, C. Caparros, R. Gonçalves, P.M. Martins, M. Benelmekki, G. Botelho, and S. Lanceros-Mendez, "Role of Nanoparticle Surface Charge on the Nucleation of the Electroactive β-Poly(vinylidene fluoride) Nanocomposites for Sensor and Actuator Applications," J. Phys. Chem. C **116**(29), 15790–15794 (2012).

[48] E. Kar, P. Ghosh, S. Pratihar, M. Tavakoli, and S. Sen, "$SiO_2$ Nanoparticles Incorporated Poly(Vinylidene) Fluoride Composite for Efficient Piezoelectric Energy Harvesting and Dual-Mode Sensing," Energy Technol. **11**(2), 2201143 (2023).

[49] J. Gil-Rostra, J. Castillo-Seoane, Q. Guo, A.B. Jorge Sobrido, A.R. González-Elipe, and A. Borrás, "Photoelectrochemical Water Splitting with $ITO/WO_3/BiVO_4$/CoPi Multishell Nanotubes Enabled by a Vacuum and Plasma Soft-Template Synthesis," ACS Appl. Mater. Interfaces **15**(7), 9250–9262 (2023).

[50] X. García-Casas, A. Ghaffarinejad, F.J. Aparicio, J. Castillo-Seoane, C. López-Santos, J.P. Espinós, J. Cotrino, J.R. Sánchez-Valencia, Á. Barranco, and A. Borrás, "Plasma engineering of microstructured piezo–Triboelectric hybrid nanogenerators for wide bandwidth vibration energy harvesting," Nano Energy **91**, 106673 (2022).

[51] J. Castillo-Seoane, L. Contreras-Bernal, T.C. Rojas, J.P. Espinós, A.-F. Castro-Méndez, J.-P. Correa-Baena, A. Barranco, J.R. Sanchez-Valencia, and A. Borras, "Highly Stable

Photoluminescence in Vacuum-Processed Halide Perovskite Core–Shell 1D Nanostructures," Adv. Funct. Mater. **34**(40), 2403763 (2024).

[52] M. Alcaire, C. Lopez-Santos, F.J. Aparicio, J.R. Sanchez-Valencia, J.M. Obrero, Z. Saghi, V.J. Rico, G. de la Fuente, A.R. Gonzalez-Elipe, A. Barranco, and A. Borras, "3D Organic Nanofabrics: Plasma-Assisted Synthesis and Antifreezing Behavior of Superhydrophobic and Lubricant-Infused Slippery Surfaces," Langmuir **35**(51), 16876–16885 (2019).

[53] N. Filippin, J. Castillo-Seoane, M.C. López-Santos, C.T. Rojas, K. Ostrikov, A. Barranco, J.R. Sánchez-Valencia, and A. Borrás, "Plasma-Enabled Amorphous $TiO_2$ Nanotubes as Hydrophobic Support for Molecular Sensing by SERS," ACS Appl. Mater. Interfaces **12**(45), 50721–50733 (2020).

[54] L. Montes, J.M. Román, X. García-Casas, J. Castillo-Seoane, J.R. Sánchez-Valencia, Á. Barranco, C. López-Santos, and A. Borrás, "Plasma-Assisted Deposition of $TiO_2$ 3D Nanomembranes: Selective Wetting, Superomniphobicity, and Self-Cleaning," Adv. Mater. Interfaces **8**(21), 2100767 (2021).

[55] K. Tashiro, H. Yamamoto, S. Kummara, D. Tahara, K. Aoyama, and H. Sekiguchi, "Electric-Field-Induced Phase Transition and Crystal Structural Change of the Oriented Poly(vinylidene Fluoride) β Form as Clarified by the In Situ Synchrotron Wide-Angle X-ray Diffraction Measurement," Macromolecules **55**(15), 6644–6660 (2022).

[56] G.T. Davis, J.E. McKinney, M.G. Broadhurst, and S.C. Roth, "Electric-field-induced phase changes in poly(vinylidene fluoride)," J. Appl. Phys. **49**(10), 4998–5002 (1978).

[57] P. Martins, A.C. Lopes, and S. Lanceros-Mendez, "Electroactive phases of poly(vinylidene fluoride): Determination, processing and applications," Prog. Polym. Sci. **39**(4), 683–706 (2014).

[58] D.W. Jin, Y.J. Ko, C.W. Ahn, S. Hur, T.K. Lee, D.G. Jeong, M. Lee, C.-Y. Kang, and J.H. Jung, "Polarization- and Electrode-Optimized Polyvinylidene Fluoride Films for Harsh Environmental Piezoelectric Nanogenerator Applications," Small **17**(14), 2007289 (2021).

[59] H.K. Mishra, A. Jain, D. Saini, B. Mondal, C. Bera, S. Ram, and D. Mandal, "Ultrahigh pyroelectricity in monoelemental two-dimensional tellurium," Phys. Rev. B **111**(15), 155436 (2025).

[60] Y. Yang, J.H. Jung, B.K. Yun, F. Zhang, K.C. Pradel, W. Guo, and Z.L. Wang, "Flexible Pyroelectric Nanogenerators using a Composite Structure of Lead-Free $KNbO_3$ Nanowires," Adv. Mater. **24**(39), 5357–5362 (2012).

[61] B. Mahanty, S. Kumar Ghosh, K. Maity, K. Roy, S. Sarkar, and D. Mandal, "All-fiber pyro- and piezo-electric nanogenerator for IoT based self-powered health-care monitoring," Mater. Adv. **2**(13), 4370–4379 (2021).

[62] J. Castillo-Seoane, J. Gil-Rostra, V. López-Flores, G. Lozano, F.J. Ferrer, J.P. Espinós, K. (Ken) Ostrikov, F. Yubero, A.R. González-Elipe, Á. Barranco, J.R. Sánchez-Valencia, and A. Borrás, "One-reactor vacuum and plasma synthesis of transparent conducting oxide nanotubes and nanotrees: from single wire conductivity to ultra-broadband perfect absorbers in the NIR," Nanoscale **13**(32), 13882–13895 (2021).

[63] A. Borras, A. Barranco, and A.R. González-Elipe, "Reversible Superhydrophobic to Superhydrophilic Conversion of $Ag@TiO_2$ Composite Nanofiber Surfaces," Langmuir **24**(15), 8021–8026 (2008).

[64] C. Ribeiro, C.M. Costa, D.M. Correia, J. Nunes-Pereira, J. Oliveira, P. Martins, R. Gonçalves, V.F. Cardoso, and S. Lanceros-Méndez, "Electroactive poly(vinylidene fluoride)-based structures for advanced applications," Nat. Protoc. **13**(4), 681–704 (2018).

[65] X. Garcia-Casas, G.P. Moreno-Martinez, F. Nuñez-Galvez, T. Czermak-Álvarez, H. Krishna, F.J. Aparicio-Rebollo, J.R. Sanchez-Valencia, A. Barranco, and A. Borras, "NanoDataLyzer," (2024).